\documentclass{article}
\pdfoutput=1
\usepackage[utf8]{inputenc}
\usepackage{authblk}
\usepackage{graphicx}
\usepackage[square,numbers,sort&compress]{natbib}
\title{Fiber to Chip Fusion Splicing for Robust, Low Loss Photonic Packaging}
\author[1]{Juniyali Nauriyal}
\author[2]{Meiting Song}
\author[2]{Raymond Yu}
\author[2,*]{Jaime Cardenas}
\affil[1]{Department of Electrical and Computer Engineering, University of Rochester, Rochester, N.Y. 14627, USA}
\affil[2]{The Institute of Optics, University of Rochester, Rochester, N.Y. 14627, USA}
\affil[*]{Corresponding author: jaime.cardenas@rochester.edu}
\date{October 2018}
\begin{document}
\maketitle
\begin{abstract}
Silicon photonic devices are poised to enter high volume markets such as data-communications, telecommunications, biological sensing, and optical phased arrays; however, permanently attaching a fiber to the photonic chip with high optical efficiency remains a challenge. We present a robust and low-loss packaging technique of permanent optical edge coupling between a fiber and a chip using fusion splicing which is low-cost and scalable to high volume manufacturing. We fuse a SMF-28 cleaved fiber to the chip via CO\textsubscript2 laser and reinforce it with optical adhesive. We demonstrate minimum loss of 1.0dB per-facet with 0.6dB penalty over 160nm bandwidth from 1480nm-1640nm.
\end{abstract}
\section{Introduction}
Silicon photonic devices are poised to enter high volume markets such as data-communications, telecommunications, biological sensing, and optical phased arrays; however, permanently attaching a fiber to the photonic chip with high optical efficiency remains a challenge \citep{kopp_silicon_2011,pavesi_optical_2007,reed_optical_2004,soref_past_2006}. Silicon photonics leverage the mature electronics fabrication infrastructure because of its compatibility with complementary metal oxide semiconductor (CMOS) fabrication. During the past few decades, the basic building blocks of silicon photonic devices have been demonstrated: modulators, detectors, switches, filters, and lasers. One of the main challenges remaining is a packaging method to permanently attach optical fibers to the photonic chips with high optical efficiency, high speed, and maintaining compatibility with CMOS processing without introducing changes to the fabrication process or consuming significant area on the chip \citep{kopp_silicon_2011,soganci_flip-chip_2013,barwicz_novel_2016,chen_low-loss_2010,pavarelli_optical_2015,snyder_packaging_2013}.

Techniques to package the fiber to the chip rely on bulky fixtures, metallic ferrules, or specialized fibers. Numerous methods for high efficiency coupling of light from an optical fiber to a photonic chip use gratings for coupling light from the top of the chip or waveguide nanotaper based couplers to couple light from the edges \citep{chen_low-loss_2010,snyder_packaging_2013,aalto_low-loss_2006,almeida_nanotaper_2003,bakir_low-loss_2010,barkai_double-stage_2008,barwicz_o-band_2015,cardenas_high_2014,cheben_refractive_2010,fang_suspended_2010,ferdous_spectral_2011,galan_polarization_2007,han_large-scale_2015,hauffe_methods_2001,kasaya_simple_1993,khilo_efficient_2010,kopp_silicon_2011,laere_compact_2007,lai_efficient_2017,masanovic_high_2005,mcnab_ultra-low_2003,mekis_grating-coupler-enabled_2011,orobtchouk_high-efficiency_2000,papes_fiber-chip_2016,pavarelli_optical_2015,roelkens_high_2006,roelkens_efficient_2005,shani_efficient_1989,shiraishi_silicon-based_2007,shoji_low_2002,shu_efficient_2011,soganci_flip-chip_2013,taillaert_compact_2004,taillaert_out--plane_2002,takei_ultranarrow_2012,vivien_light_2006,wang_low-loss_2016,wood_compact_2012,xia_asymmetric_2001,yoda_two-port_2009,zengerle_low-loss_1992,zhu_very_2017}. While grating couplers have larger alignment tolerance and give optical access from the top of the chip, they have a narrow bandwidth and typically require bulky fixtures to attach them to the chip \citep{kwong_-chip_2014,laere_compact_2007,masanovic_high_2005,mekis_grating-coupler-enabled_2011,orobtchouk_high-efficiency_2000,taillaert_compact_2004,taillaert_out--plane_2002,vivien_light_2006}. Edge couplers are broadband but have a tighter alignment tolerance. They typically use bulky fixtures, high NA fibers, or lensed fibers to attach them to the chip. In these techniques the fibers are usually permanently attached to the chip using optical adhesives \citep{bakir_low-loss_2010,han_large-scale_2015,hauffe_methods_2001,kopp_silicon_2011,lai_efficient_2017,mekis_grating-coupler-enabled_2011,papes_fiber-chip_2016,wang_low-loss_2016,almeida_nanotaper_2003}. Fixtures and lensed fibers significantly increase the cost of the packaging.
\begin{figure}[htbp]
\centering
\includegraphics[width=\linewidth]{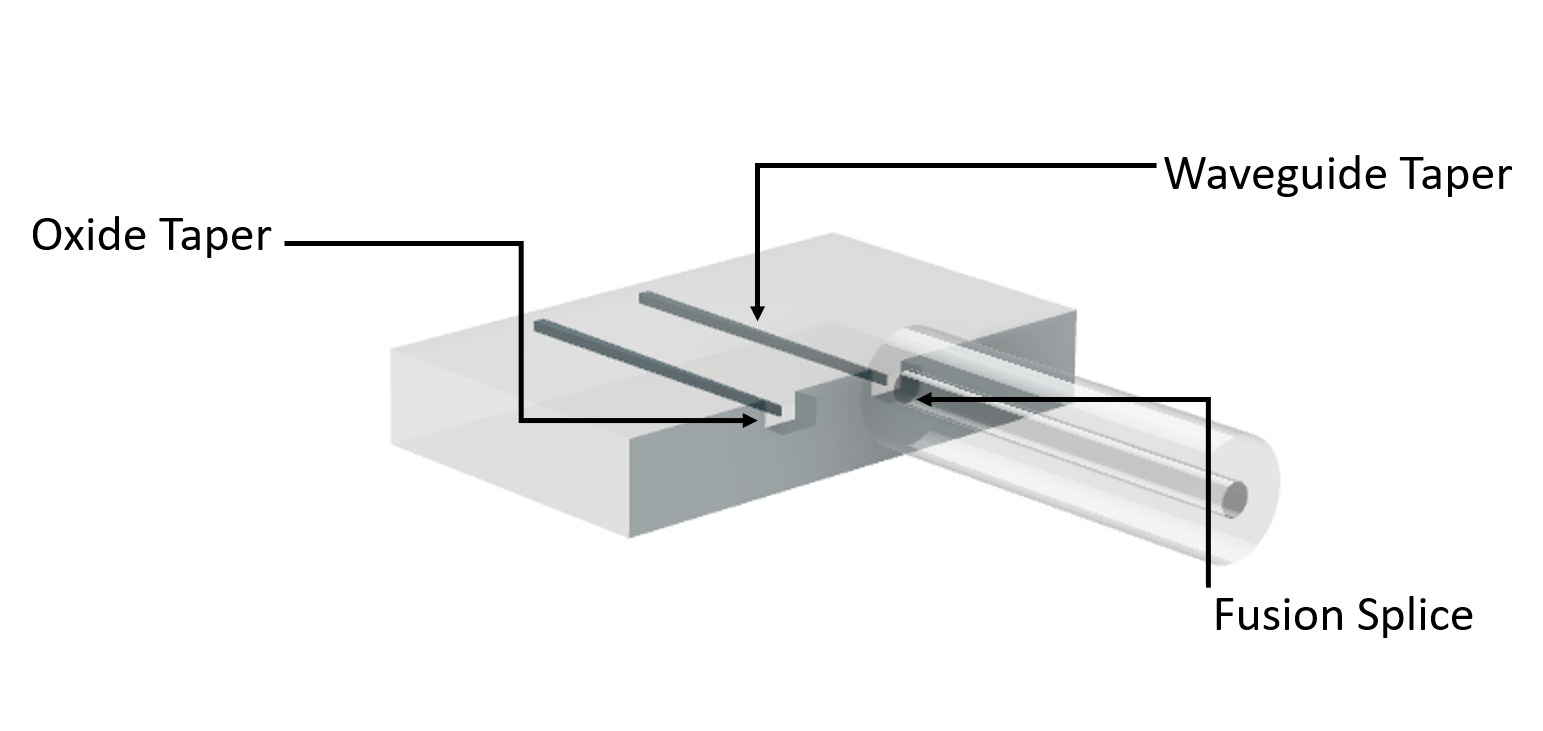}
\caption{3D model of fiber to chip packaging using fusion splicing (not to scale).}
\label{fig:3d}
\end{figure}
\section{Methods}
We present a low-cost, robust and low-loss packaging technique of permanent optical edge coupling between a fiber and a chip using fusion splicing which is scalable to high volume manufacturing and has a larger alignment tolerance. Our approach consists of a cantilever-type silicon dioxide waveguide \citep{chen_low-loss_2010,wood_compact_2012}, which is mode-matched to a single mode fiber on one side and to a waveguide nanotaper on the other.
\begin{figure}[htbp]
\centering
\includegraphics[width=\linewidth]{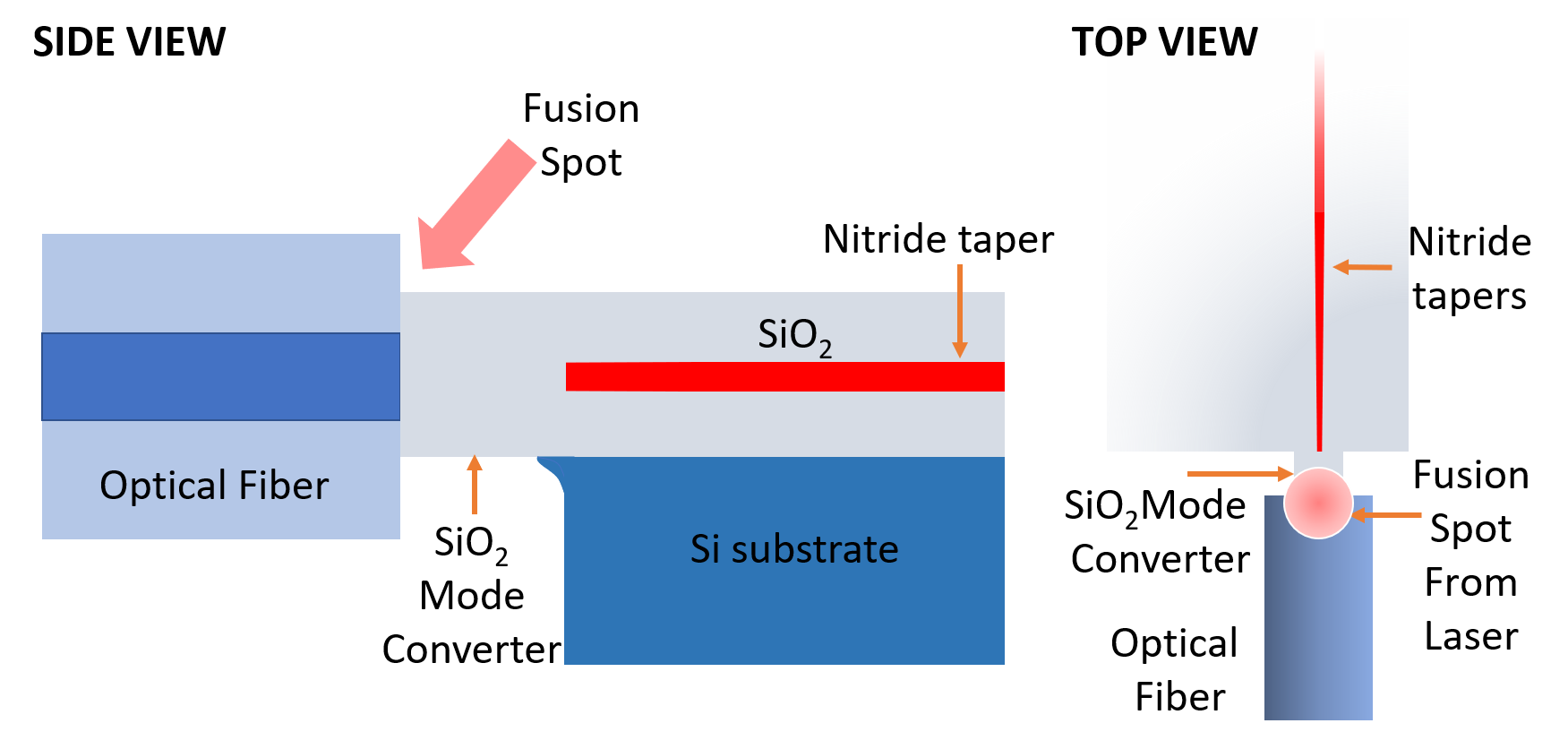}
\caption{Schematic representation of a packaged device using silicon dioxide mode converter fused to SMF-28 fiber. The side view shows undercut silicon substrate which isolates the oxide mode converter from the chip, and the top view of the method shows the spot where the fiber and the chip are fused to improve coupling and increases mechanical stability.}
\label{fig:topview}
\end{figure}

The oxide waveguide is permanently fused to the optical fiber (figure \ref{fig:3d}) using a CO\textsubscript2 laser via radiative energy. The fusion splice between the fiber and the chip forms a permanent bond and decreases coupling losses by eliminating the Fresnel reflections at both oxide-air interfaces and the gap between the fiber and the oxide taper. This method is compatible with different types of inverse nanotapers (e.g. linear taper, metamaterial taper)  \citep{almeida_nanotaper_2003,rowland_tapered_1991,kippenberg_microresonator-based_2011} since the oxide waveguide geometry can be engineered to match the nanotaper mode profile. The oxide waveguide is carved out from the upper and under claddings of the device. It is compatible with standard foundry processes and does not require adding or removing steps from the typical fabrication process. The proposed method does not require special blocks or fiber holders to hold the fibers and uses standard, cleaved optical fibers. 
\begin{figure}[!htbp]
\centering
\includegraphics[width=\linewidth]{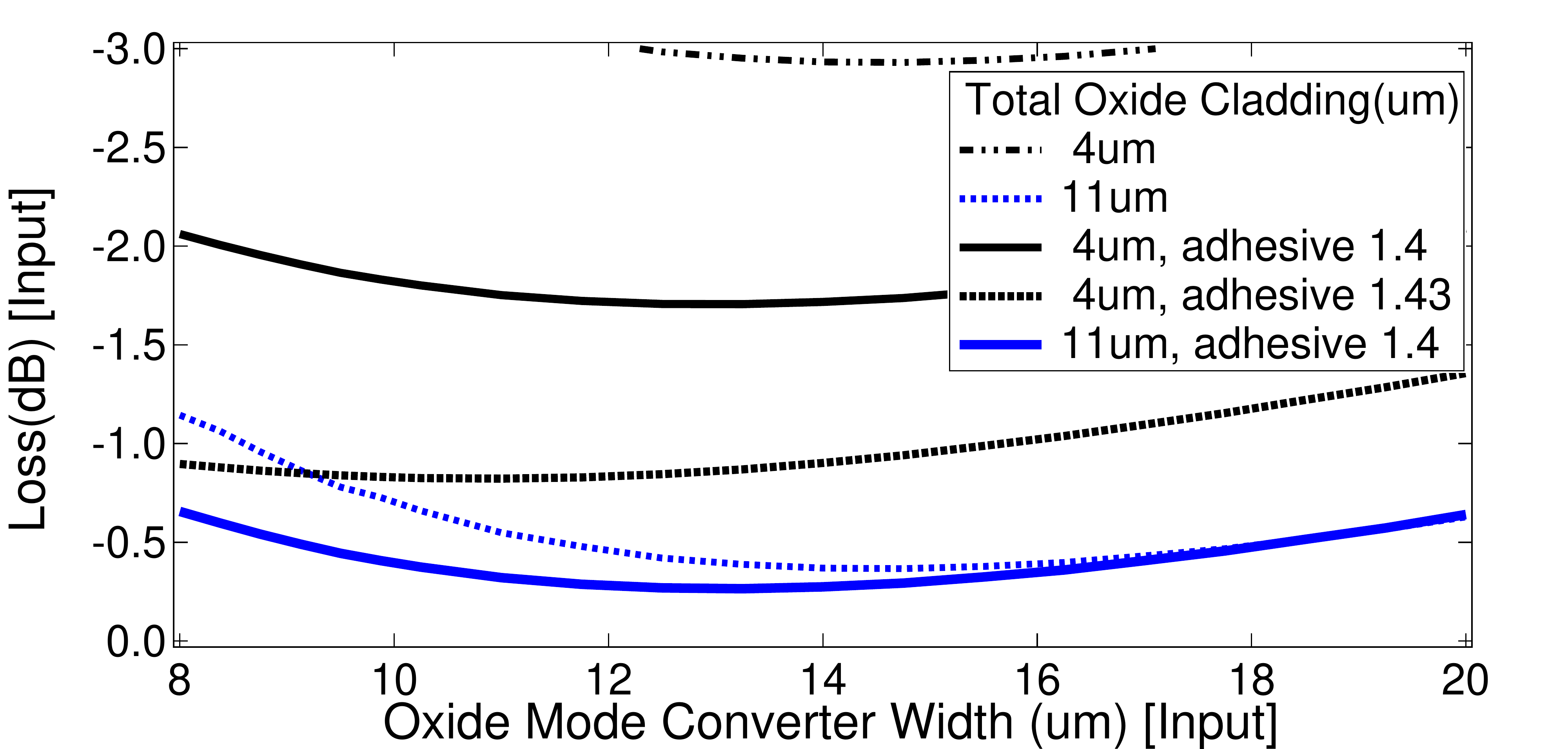}
\caption{Coupling loss between the fiber and the mode converter as a function of oxide mode converter input width. A minimum coupling loss of 0.3dB can be achieved with 14$\mu$m oxide mode converter width and 11$\mu$m total (top plus bottom) oxide cladding thickness with an adhesive of refractive index 1.4.}
\label{fig:clad}
\end{figure}
The silicon dioxide mode converter is designed to efficiently couple to a cleaved optical fiber on one of its ends and to a waveguide nanotaper on the other end (figure \ref{fig:topview}). The mode conversion process occurs in two stages: from the waveguide (mode size $<$1$\mu$m) to the waveguide nanotaper and then from the oxide mode converter to the optical fiber (mode size of 10.4$\mu$m). The geometry of the oxide mode converter is engineered to optimize coupling by matching the modes of the optical fiber with the oxide mode converter. We adiabatically vary the oxide mode converter to maximize the coupling into the waveguide taper. The optimum dimensions of the oxide mode converter widths depend mainly on the refractive index of the guiding medium and the thickness of oxide cladding. This coupling method can be used in fusing any device that has a cladding of silicon dioxide including devices based on silicon, silicon nitride, and lithium niobate on insulator, for example. We isolate the oxide mode converter from the silicon substrate to prevent loss of light to the substrate. Due to the physical size mismatch between the fiber and the oxide mode converter, we reinforce the splice with a UV curable optical adhesive. Controlling the refractive index of the optical adhesive will further tailor the properties of the oxide taper mode and improve the coupling to the fiber \citep{chen_low-loss_2010}.
\begin{figure}[htbp]
\centering
\includegraphics[width=\linewidth]{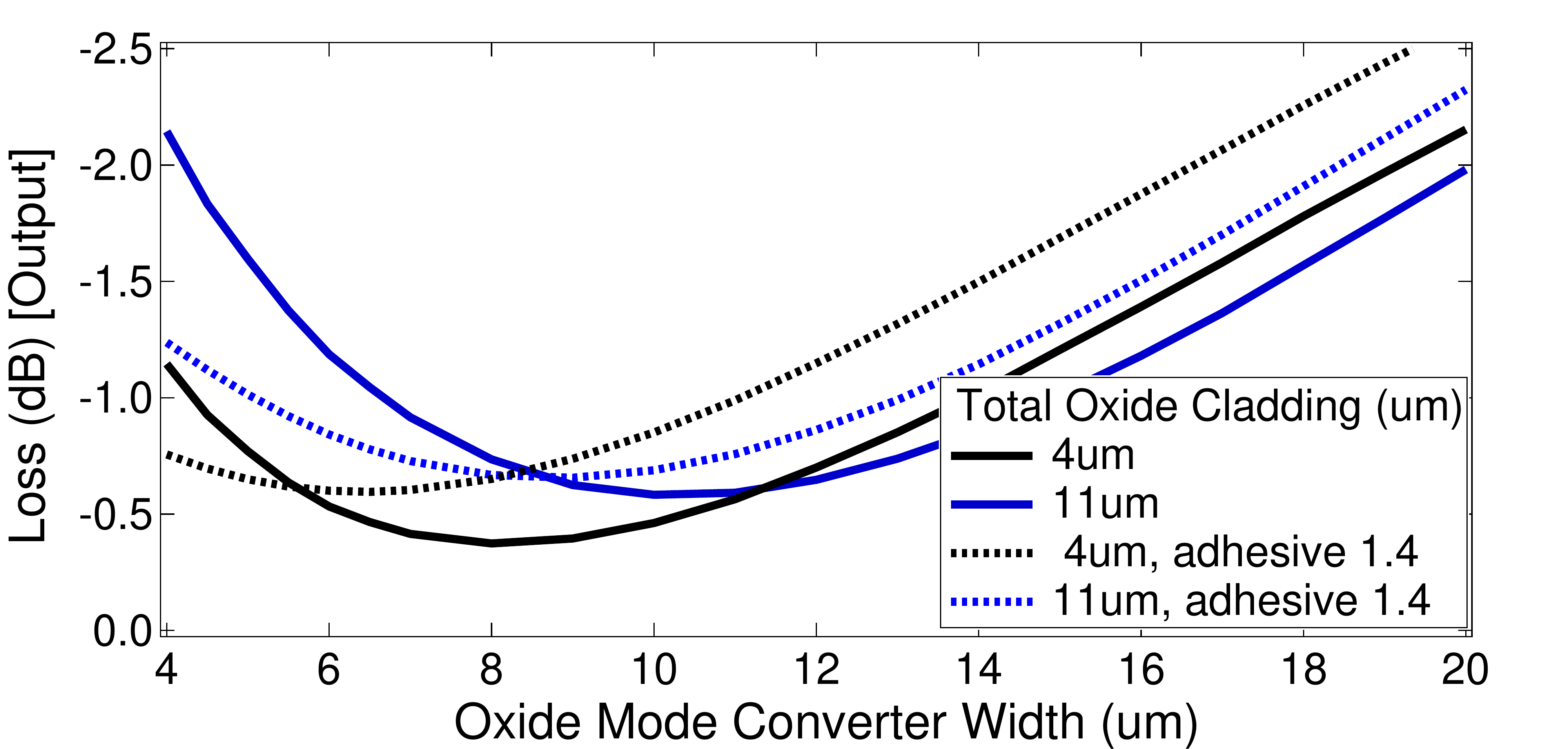}
\caption{Coupling loss between the oxide mode converter and the waveguide nanotaper as a function of output mode converter width. Loss is sensitive to mode converter width and only weakly dependent on total cladding thickness.}
\label{fig:c2}
\end{figure}
\begin{figure}[!htbp]
\centering
\includegraphics[width=\linewidth]{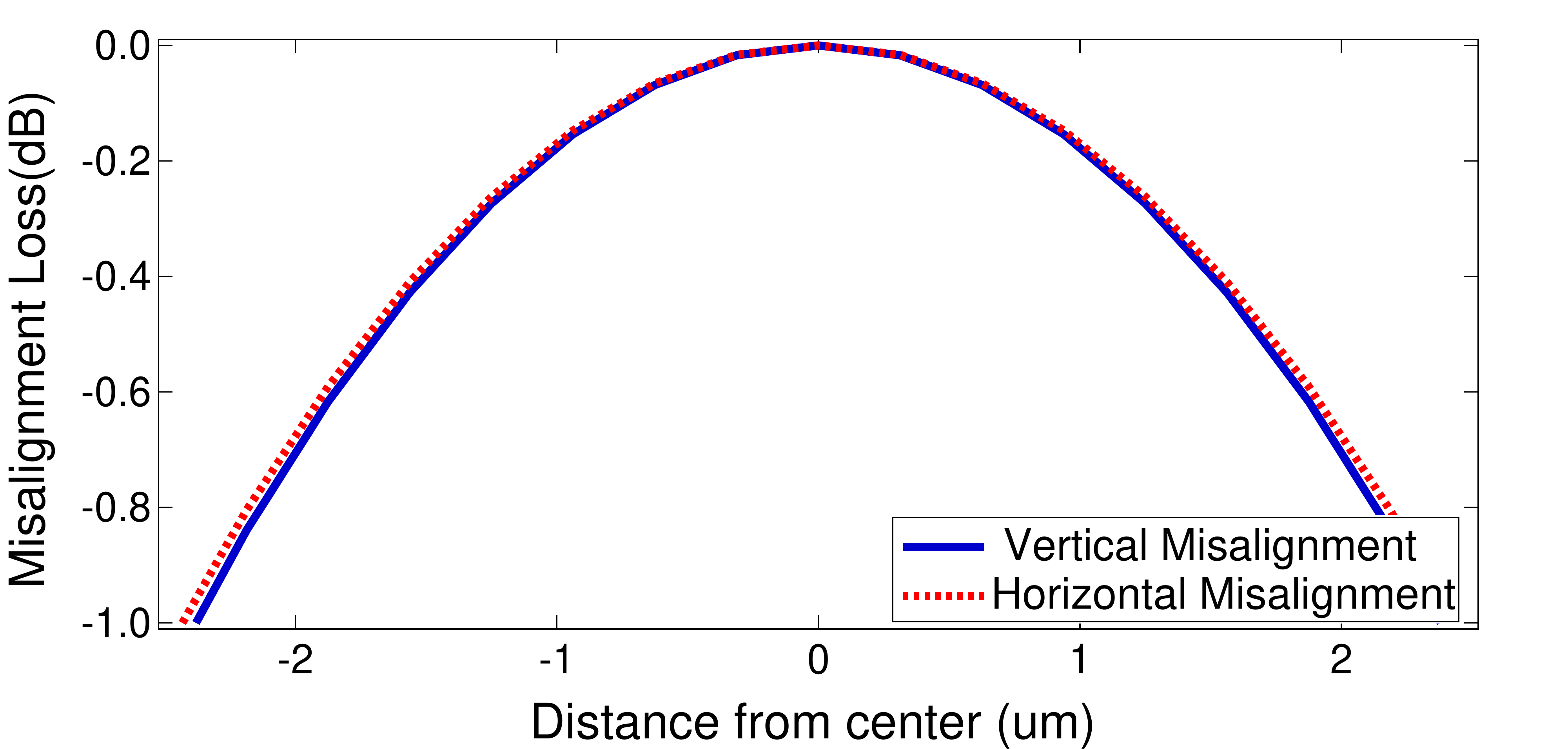}
\caption{Coupling loss due to misalignment between fiber and oxide mode converter. The tolerance for 1dB loss is +/- 2.5$\mu$m and +/- 2.4$\mu$m in the horizontal and vertical direction respectively.}
\label{fig:mis}
\end{figure}

We simulate the coupling loss for TE polarization between a silicon nitride waveguide nanotaper and the optical fiber using the eigenmode expansion method (FIMMPROP by Photon Design). The waveguide nanotaper is 0.18$\mu$m wide at the tip, 100$\mu$m long, and has a linear profile. We calculate the coupling loss between the optical fiber and the input of the oxide mode converter by launching a gaussian mode with a 10.4$\mu$m diameter into the mode converter. Our calculations (figure \ref{fig:clad}) show that coupling improves as the oxide cladding thickness increases for an oxide mode converter width of between 13$\mu$m and 15$\mu$m. Using the optical properties of the adhesive used to stabilize the splice can further tailor the mode matching and enhances the coupling by expanding the mode even in cases with thinner cladding thicknesses \citep{chen_low-loss_2010}. The losses at the output of the mode converter are calculated by launching the fundamental mode of the input oxide mode converter waveguide (figure \ref{fig:c2}). The coupling loss at the oxide mode converter-nanotaper interface is dominated by the nanotaper design and is weakly dependent on the oxide cladding thickness. The 1dB penalty misalignment tolerance between the fiber and the oxide mode converter is +/- 2.5$\mu$m and +/-2.4$\mu$m in the horizontal and vertical directions, respectively (figure \ref{fig:mis}); which compares favorably with other edge coupling methods \citep{wang_low-loss_2016,papes_fiber-chip_2016,lai_efficient_2017,kopp_silicon_2011,hauffe_methods_2001,han_large-scale_2015,bakir_low-loss_2010}.

We fabricate a silicon nitride photonic chip using standard, CMOS compatible, microfabrication techniques. A 5.5$\mu$m layer of silicon dioxide is deposited via plasma enhanced chemical vapor deposition (PECVD) and 325nm of silicon nitride are deposited via low-pressure chemical vapor deposition (LPCVD). The waveguides are patterned using standard DUV optical lithography at 254nm using an ASML stepper. After etching in an inductively coupled plasma reactive ion etcher (ICP-RIE) with a CHF\textsubscript3/O\textsubscript2 chemistry, the devices are clad with 5.8$\mu$m of oxide using plasma enhanced chemical vapor deposition (PECVD). We then pattern and etch the oxide mode converter similarly to the waveguide step. After dicing, we remove the silicon substrate using xenon difluoride (XeF\textsubscript2) dry etch to optically isolate the oxide mode converter from the silicon substrate.

We fuse a SMF 28 cleaved fiber with a mode field diameter of 10.4$\mu$m to the photonic chip by using a CO\textsubscript2 laser and reinforce the splice with an optical adhesive. Silicon dioxide strongly absorbs light at a 10.6$\mu$m wavelength from the CO\textsubscript2 laser. The laser radiatively heats the silicon dioxide fiber as well as the cladding of the chip. Radiative heating for splicing leaves no residue behind \citep{pal_low_2008,shimizu_fusion_nodate,egashira_optical_1977}. The laser beam from the CO\textsubscript2 laser is focused to a spot of 45$\mu$m using a ZnSe aspheric lens (f=20mm) and aimed at the fiber-chip interface with a co-linear red diode laser for alignment. The laser beam is incident at an angle of 30$^\circ$, which enables fusion of multiple fibers to a single device. To fuse the fiber to the oxide mode converter, we irradiate the spot with 9W of laser power for 0.5 seconds.
\section{Results}
\begin{figure}[htbp]
\centering
\includegraphics[width=\linewidth]{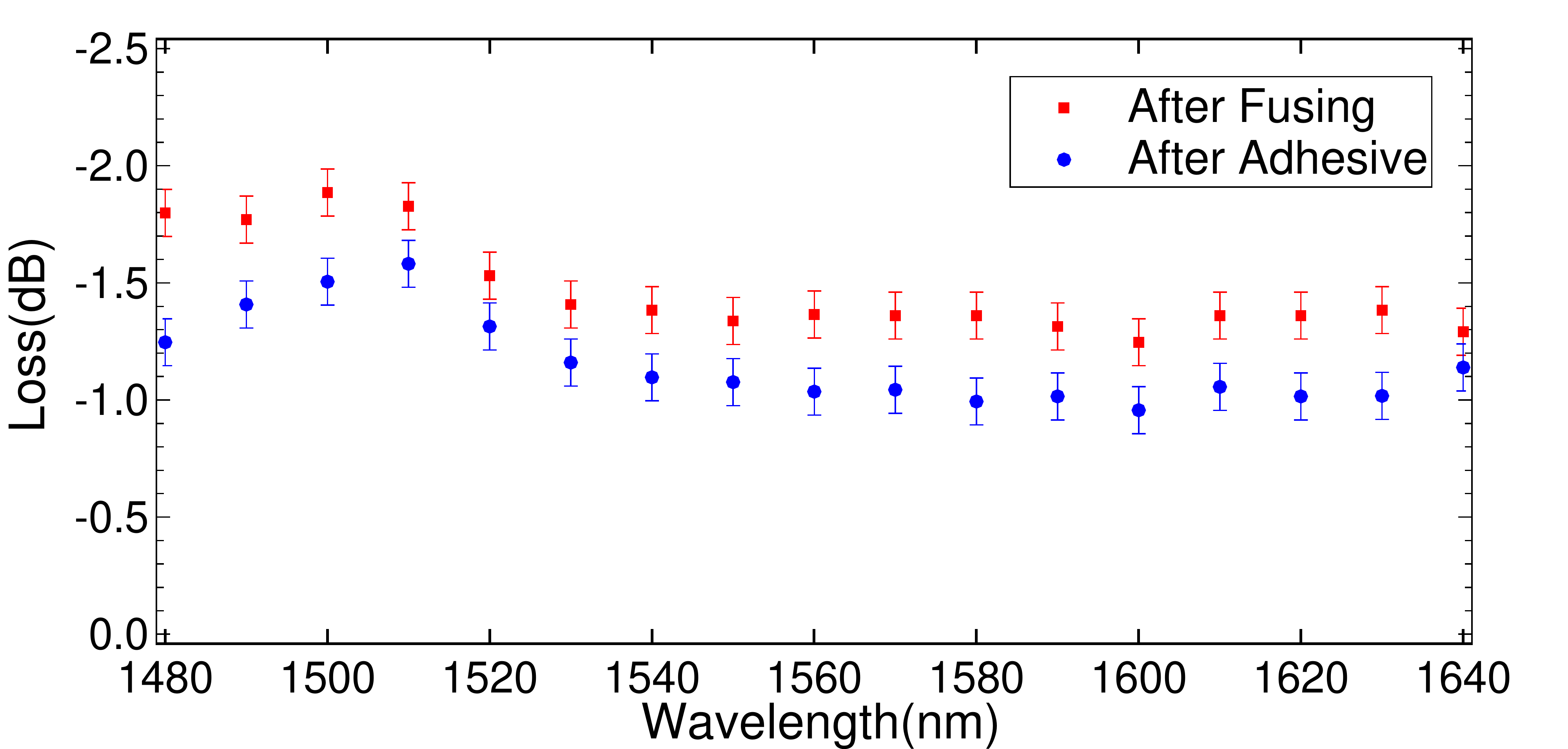}
\caption{Coupling loss per facet as a function of wavelength. The measured loss is 1.0dB per facet with a 0.6dB penalty over 160nm bandwidth.}
\label{fig:result}
\end{figure}

	We demonstrate a minimum loss of 1.0dB per facet with a 0.6dB penalty over 160nm bandwidth near the C-band on a standard, cleaved SMF-28 fiber fused to the silicon nitride photonic chip. To measure the coupling loss, we first measure the optical power exiting the input fiber. We align cleaved fibers to the oxide mode converters at the input and output of the chip and measure a loss of 2.1dB per facet (the total loss is 4.6dB and both coupling regions are the same within fabrication variation and 0.4dB are waveguide propagation losses.) We measure the loss of 0.4dB through the nitride waveguide by collecting the output with a microscope objective in place of the output cleaved fiber and comparing the two measurements.  After fusing the fiber to the mode converter on the input side, we measure a total loss of 3.8dB. We subtract the 2.1dB loss from the output fiber and the 0.4dB loss from the waveguide to find the coupling loss of the fused fiber of 1.3dB (figure \ref{fig:result}). The loss is reduced after splicing because Fresnel reflections are eliminated (approximately 0.3dB of loss) and the small gap between fiber and mode converter disappears. We apply an optical adhesive with a refractive index of 1.3825, as specified by the manufacturer, to stabilize the splice. The optical adhesive expands the mode and reduces the coupling loss to 1.0dB.
\section{Discussion}
 The total coupling loss is the sum of the losses from the fiber to oxide mode converter coupling and the oxide mode converter to waveguide nanotaper (loss in figure \ref{fig:clad} plus loss in figure \ref{fig:c2}). The loss at the fiber to oxide mode converter can be minimized by increasing the oxide cladding thickness or by optimizing the refractive index of the optical adhesive. In general both methods are used together to minimize coupling loss; however, in cases when the cladding thickness can’t be changed, such as in silicon on insulator based devices where the buried oxide is typically 3$\mu$m or less, choosing the optimum optical adhesive is critical for achieving low loss. The loss between the oxide mode converter and the waveguide nanotaper can be minimized through a suitable choice of nanotaper tip width, taper geometry, and taper type (e.g. continuous vs. metamaterial). The packaging method described here is flexible to accommodate different design choices depending on the photonic platform being considered. It is compatible with any platform that uses silicon dioxide as a cladding material.
\section{Conclusions}
Fiber to chip fusion splicing has the potential to enable high throughput optical packaging with a robust, high efficiency, and low-cost solution. The method is compatible with multiple photonic platforms, such as silicon, silicon nitride, and lithium niobate, that use silicon dioxide as the cladding. We envision that this method can be fully automated to enable highly efficient fiber to chip coupling in high volume applications and can be extended to passive alignment techniques.
\section*{Funding Information}
Hajim School of Engineering and Applied Sciences, University of Rochester

\section*{Acknowledgments}
This work was performed in part at Cornell NanoScale Facility, an NNCI member supported by NSF Grant ECCS-1542081. 

\bibliographystyle{unsrtnat}
\end{document}